\documentclass[journal=apchd5,manuscript=article]{achemso}

\usepackage{chemformula} 
\usepackage[T1]{fontenc}
\usepackage{siunitx} 
\usepackage{booktabs}
\usepackage{caption}
\usepackage{scalerel}
\usepackage{float}
\usepackage{amssymb}
\usepackage{natbib}
\usepackage{hyperref}

\author{Sean Lynch}
\affiliation{Miner School of Computer Science\\
	University of Massachusetts Lowell\\
	Lowell, MA 01854, USA}
\author{Jacob LaMountain}
\email{jacob_lamountain@student.uml.edu}
\author{Bo Fan}
\affiliation{Department of Physics and Applied Physics\\
	University of Massachusetts Lowell\\
	Lowell, MA 01854, USA}
\author{Jie Bu}
\affiliation{Department of Computer Science\\
	Virginia Tech\\
	Blacksburg, VA 24061, USA}
\author{Amogh Raju}
\author{Daniel Wasserman}
\affiliation{Department of Electrical and Computer Engineering\\
	University of Texas Austin\\
	Austin, TX 78712, USA}
\author{Anuj Karpatne}
\affiliation{Department of Computer Science\\
	Virginia Tech\\
	Blacksburg, VA 24061, USA}
\author{Viktor A. Podolskiy}
\affiliation{Department of Physics and Applied Physics\\
	University of Massachusetts Lowell\\
	Lowell, MA 01854, USA}

\title{Physics-guided hierarchical neural networks for Maxwell's equations in plasmonic metamaterials}

\begin{document}
	\onehalfspacing
	\clearpage
	
	\begin{abstract}
		
		While machine learning (ML) has found multiple applications in photonics,   traditional ``black box'' ML models typically require prohibitively large training data sets. Generation of such data, as well as the training processes themselves, consume significant resources, often limiting practical applications of ML. Here we demonstrate that embedding Maxwell's equations into ML design and training significantly reduces the required amount of data and improves the physics-consistency and generalizability of ML models, opening the road to practical ML tools that do not need extremely large training sets. The proposed physics-guided machine learning (PGML) approach is illustrated on the example of predicting complex field distributions within hyperbolic metamaterial photonic funnels, based on multilayered plasmonic-dielectric composites. The hierarchical network design used in this study enables knowledge transfer and points to the emergence of effective medium theories within neural networks. \end{abstract}

	\newpage
	\section{Introduction}
	Composite materials with engineered optical properties,  metamaterials and metasurfaces, are rapidly advancing as platforms for optical communications, sensing, imaging, and computing \cite{vpTutorialsBook, vpBoltassevaMetasurface, vpBrongersmaMetasurface, vpEnghetaBook,vpEleftheriadesMetasurface, eager-1, eager-2}. The complexity of typical metamaterials makes it almost impossible to understand and optimize their interaction with light based on experimental or analytical theory approaches alone, leaving the problem of light interaction with metamaterials to computational sciences\cite{vpTutorialsBook,vpCapassoMetasurface,vpEnghetaBook, Banks2019}. Currently, finite-difference time domain (FDTD) \cite{vpFTDTbook} and finite element methods (FEM) \cite{vpFEMbook,comsol} represent industry-standard approaches to understanding the optics of nonperiodic composite media.   
	
	Machine learning (ML) techniques, particularly neural networks (NNs), have recently been incorporated into the design, evaluation, and measurement of nanophotonic structures\cite{vpMLSuchowski, Khatib2021, Jiang2021, lin2021end, kudyshev2023machine, kudyshev2021optimizing,li2024machine,pestourie2020active,zhelyeznyakov2023large}. Properly trained ML tools can be used as surrogate models that predict the spectral response of composites or---rarely---field distributions within metamaterials\cite{Khatib2021, vpMLLiu, vpMLLiu02, lu2021physics}. Since ML does not solve the underlying electromagnetic problem, these predictions are  significantly faster than brute-force simulations. However, extensive training sets, often featuring $\sim~10^3\ldots 10^5$ configurations are required in order to develop high-quality ML models\cite{hestness2019,zhao2020roledataset,jiang2019global,mao2024towards,Khatib2021}. The time and computational resources needed to generate these data sets, as well as the time and resources needed for the ML training process, are significant and often serve as the main limitation to ML use in computational photonics.  
	
	Embedding physics-based constraints (physics-consistency)\cite{physicsConsistency1,physicsConsistency2} into the ML training process may be beneficial for the resulting models. ML methods for general solutions of partial differential equations (PDEs) are being developed\cite{BS2019, wang2021learning, lu2021learning, seidman2022nomad, goswami2022physics, li2020fourier, Yao2020EnhancedDL, raissi2019physics}. However, as of now, these techniques are illustrated on convenient ``toy'' models and cannot be straightforwardly applied to practical electromagnetic problems. 
	Physics-guided machine learning (PGML)\cite{eager-1,eager-2,chen2022high} is emerging as a promising platform that can combine data- and physics-driven learning. Notably, previous PGML attempts have been focused on dielectric\cite{chen2022high} or relatively simple plasmonic\cite{eager-2} composites. Here we develop PGML models that are capable of predicting electromagnetic fields within plasmonic metamaterials. We illustrate our technique by analyzing the optical response of metamaterials-based photonic funnels \cite{funnelsVP, funnels2006, funnels2024}: conical structures with strongly anisotropic composite cores that are capable of concentrating light to deep subwavelength areas. We show that physics-based constraints enable training on unlabeled data and significantly improve the accuracy and generalizability of the models. We also attempt to understand the inner workings of the NNs by analyzing the performance of hierarchical models with different data resolutions.  
	
	\begin{figure}[h!]
		\centering
		\includegraphics[width=0.8\textwidth]{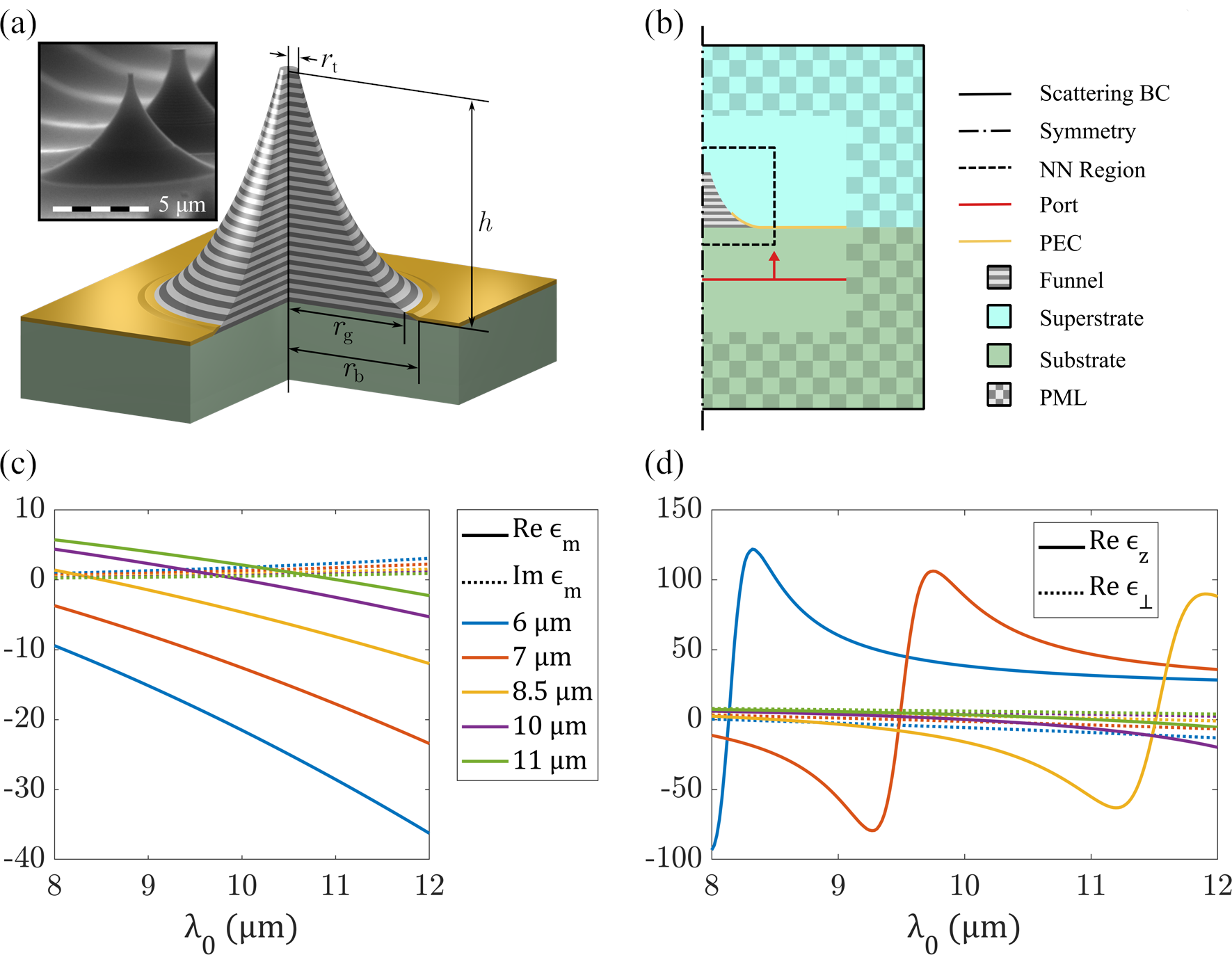}
		\caption{(a) schematic of the photonic funnel with cut-out demonstrating the composite structure of the core; inset shows scanning electron microscope (SEM) image of as-fabricated array of funnels (b) simulation setup used in FEM-based solutions of Maxwell's equations; NNs are trained only on a subregion of the FEM data which contains the funnel;  (c,d) wavelength dependence of (c) the permittivity of the highly doped plasmonic components of the funnels and (d) the components of the resulting effective permittivity tensor.}
		\label{fig:schematic}
	\end{figure}

	\section{Hyperbolic Metamaterial-based Photonic Funnels}
	
	An electromagnetic composite comprising sufficiently thin alternating layers of non-magnetic materials with permittivities $\epsilon_1,\epsilon_2$ and thicknesses $d_1,d_2$ (see Fig.1) behaves as a uniaxial medium whose optical axis is perpendicular to the layers (direction $\hat{z}$ in this work) and whose diagonal permittivity tensor has components given by: $\epsilon_{xx}=\epsilon_{yy}=\epsilon_\perp=\frac{d_1\epsilon_1+d_2\epsilon_2}{d_1+d_2}$ and $\epsilon_{zz}=\frac{(d_1+d_2)\epsilon_1\epsilon_2}{d_1\epsilon_2+d_2\epsilon_1}$. Such a material supports the propagation of two types of plane waves that differ in their polarization and have fundamentally different dispersions. 
	
	The ordinary waves (which have $\vec{E}\perp \hat{z}$) satisfy the dispersion relation $k^2=\epsilon_\perp\omega^2/c^2$ with $\vec{k}, \omega$, and $c$ representing the wavevector of the wave, operating angular frequency, and speed of light in vacuum, respectively. This dispersion is identical to that of plane waves propagating in a homogeneous isotropic material with permittivity $\epsilon_\perp$. On the other hand, the extraordinary, or transverse-magnetic (TM), waves (with $\vec{H}\perp \hat{z}$) have dispersion ${\frac{k_x^2+k_y^2}{\epsilon_{zz}}+\frac{k_z^2}{\epsilon_\perp}=\frac{\omega^2}{c^2}}$. Notably, for anisotropic materials, the dispersion of extraordinary waves is either elliptical or hyperbolic.  The topology of the iso-frequency contours strongly depends on the combination of signs of the effective permittivity tensor. 
	
	When the components of the permittivity tensor are of opposite signs, the iso-frequency surfaces are hyperboloids. This hyperbolic dispersion has been identified as the enabling mechanism for such unique optical phenomena as negative refraction, strong enhancement of light matter interaction, and for light manipulation in ultrasmall (deep subwavelength) areas \cite{vpHyperReview, vpBoltassevaMetasurface, vpNoginovProkes, vpIndexSensingOptExp, vpNarimanovExtraction}. Hyperbolicity can be achieved by alternating layers of dielectric ($\epsilon_1=\epsilon_d>0$) and plasmonic ($\epsilon_2=\epsilon_m<0$) layers. In the semiclassical regime (typically, when the layer thickness $\gtrsim$ \SI{10}{\nano\meter}) the permittivity of the plasmonic layers as a function of angular frequency of light is well described by  the Drude model\cite{jackon},
		\begin{equation}
			\epsilon_m(\omega)=\epsilon_\infty\left(1-\frac{\omega_p^2}{\omega^2+i\gamma\omega}\right)
		\end{equation}
		with $\epsilon_\infty, \omega_p$, and $\gamma$ being background permittivity,  plasma frequency, and scattering rate, respectively. Here we use $\epsilon_\infty=12.15, \gamma=10^{13} s^{-1}$,  and parameterize the plasma frequency using the plasma wavelength, $\lambda_p$, via $\omega_p=2\pi c/\lambda_p$.
	
	The most common implementation of these metamaterials leverages a 50/50 composition $(d_1=d_2=d)$. For such systems topological transitions occur when the permittivity of the plasmonic layers ($\epsilon_m$) and the weighted permittivity of the mixture ($\epsilon_\perp$) change signs\cite{hypNat2013,shekhar2014hyperbolic}. The dispersion of TM waves inside the metamaterial is  elliptic  for shorter wavelengths $\lambda<\lambda_p$. It changes to type-I hyperbolicity ($\epsilon_\perp>0,\epsilon_{zz}<0$) for $\lambda_p<\lambda<\tilde{\lambda}_p$, with the renormalized plasma frequency, $\tilde{\lambda}_p$, defined as ${\rm Re}(\epsilon_\perp(\tilde{\lambda}_p))=0$. Finally, the dispersion of TM waves in the composite becomes type-II hyperbolic ($\epsilon_\perp<0,\epsilon_{zz}>0$) for $\tilde{\lambda}_p<\lambda$. 
	
	Photonic funnels, conical waveguides with hyperbolic metamaterial cores, \cite{funnels2006,funnelsVP,funnels2024} shown in Fig.1, represent excellent examples of structures capable of manipulating light at a deep subwavelength scale. Recent experimental results\cite{funnelsVP,funnels2024} demonstrate efficient concentration of mid-infrared light with a vacuum wavelength of \SI{\sim 10}{\micro\meter} to spatial areas as small as \SI{\sim 300}{\nano\meter}, 1/30\textsuperscript{th} of the operating wavelength, within an all-semiconductor ``designer metal'' material platform \cite{shekhar2014hyperbolic}. Further analysis relates the field concentration near the funnels' tips to the absence of the diffraction limit within the hyperbolic material and to the anomalous internal reflection of light  from the funnel sidewall, which forms an interface oblique to the optical axis.\cite{funnels2024} Importantly, the optical response of realistic funnels can be engineered at the time of fabrication by controlling the doping of the  ``designer metal'' layers and thereby adjusting the plasma frequency of these layers.  The unusual electromagnetic response, strong field confinement, and significant field inhomogeneities make photonic funnels an ideal platform for testing the performance of ML-driven surrogate solvers of Maxwell's equations.

	\section{Methods}

	\subsection{Data set description and generation}
	To construct a sufficiently diverse set of labeled configurations, we used FEM to solve for electromagnetic field distributions in photonic funnels with plasmonic layers of different doping concentrations corresponding to plasma wavelengths\cite{law2012mid} of \SIlist{6; 7; 8.5; 10; 11}{\micro\meter}. 
	Fig.\ref{fig:schematic} illustrates the wavelength-dependent permittivity of plasmonic layers with various doping concentrations as well as the corresponding effective medium response of the layered metamaterials. Note the drastic changes of effective medium response as a function of both wavelength and doping.   
	
	For each doping level, wavelength-dependent permittivity and electromagnetic field distributions have been calculated with a commercial FEM-based solver\cite{comsol} (that takes into account that all fields are proportional to $\exp(-i\phi)$ with $\phi$ being the angular coordinate of the cylindrical reference frame) for free-space wavelengths from \qtyrange{8}{12}{\micro\meter} with increments of \SI{62.5}{\nano\meter}. The FEM model setup is shown schematically in Fig.\ref{fig:schematic}a. Electromagnetic waves that are normally incident on the funnel base are generated by the port boundary condition. Perfectly matched layers\cite{vpFEMbook} and scattering boundary conditions are used to make the outside boundaries of the simulation region completely transparent to electromagnetic waves, thereby mimicking the surrounding infinite space. The model, which explicitly incorporates \SI{80}{\nano\meter}-thick layers in the funnel cores, is meshed with a resolution of at most \SI{40}{\nm} inside the funnel and \SI{200}{nm} outside the funnels, with the mesh growth factor set to 1.1 to avoid artifacts related to abrupt changes in mesh size. 
	
	For every plasma wavelength and operating frequency the distribution of electromagnetic fields, along with the distributions of permittivities within a small ($5\times12$ \unit{\micro\meter}) region of space containing the funnel (see Fig.\ref{fig:schematic}) is interpolated onto a rectangular mesh with resolution \SI{12.5}{\nano\meter} $\times$ \SI{10}{\nano\meter} along the $r$ and $z$ directions, respectively, forming the basis for the data sets used in the study. Note that selecting this internal region of space from the FEM simulations allows us to (i)  implicitly incorporate the proper boundary conditions for both incident as well as scattered electromagnetic fields  and (ii) avoid the implementation of perfectly matched layers, ports, and scattering conditions within the physics-based constraints used in training our NNs.

	The original FEM-generated data has been then resampled into three separate data sets: 
	\begin{itemize}
		\item low-resolution data set, $20\times60$ pixels with resolution \SI{250}{\nano\meter} and \SI{200}{\nano\meter} in $r$ and $z$ directions, respectively
		\item medium-resolution data set, $100\times300$ pixels with resolution \SI{50}{\nano\meter} $\times$ \SI{40}{\nano\meter}
		\item high-resolution data set, $200\times 600$ pixels with resolution of \SI{25}{\nano\meter} $\times$ \SI{20}{\nano\meter}
	\end{itemize}

	\section{Neural Network Architecture}
	
	On a fundamental level, approximating solutions of Maxwell's equations within metamaterials with ML necessitates a neural network to map the operating frequency and the distribution of permittivity across the composite to the distribution of electromagnetic fields, a problem that is similar to image transformation. Previous analysis\cite{hou2017,8456517} has demonstrated that convolutional neural networks (CNNs) excel in image transformation. Specifically, encoder-decoder, CNN, and U-net architectures have shown success in electromagnetic problems\cite{9444655,9913707,kudyshev2020machine,li2024machine,zhelyeznyakov2023large}, presumably due to the cores of the networks learning some low-dimensional representation of the solutions\cite{Gong2020,seidman2022nomad}. Note, however, that the vast majority of previous ML-driven solvers of Maxwell's equations\cite{lim2022maxwellnet,ZhangMaxwell,chen2022high,zhelyeznyakov2023large} have analyzed dielectric composites (where electromagnetic fields are relatively smooth) and were trained on relatively large data sets\cite{chen2022high,Khatib2021,pestourie2020active}. 
	
\begin{figure}
		\centering
		\includegraphics[width=\linewidth]{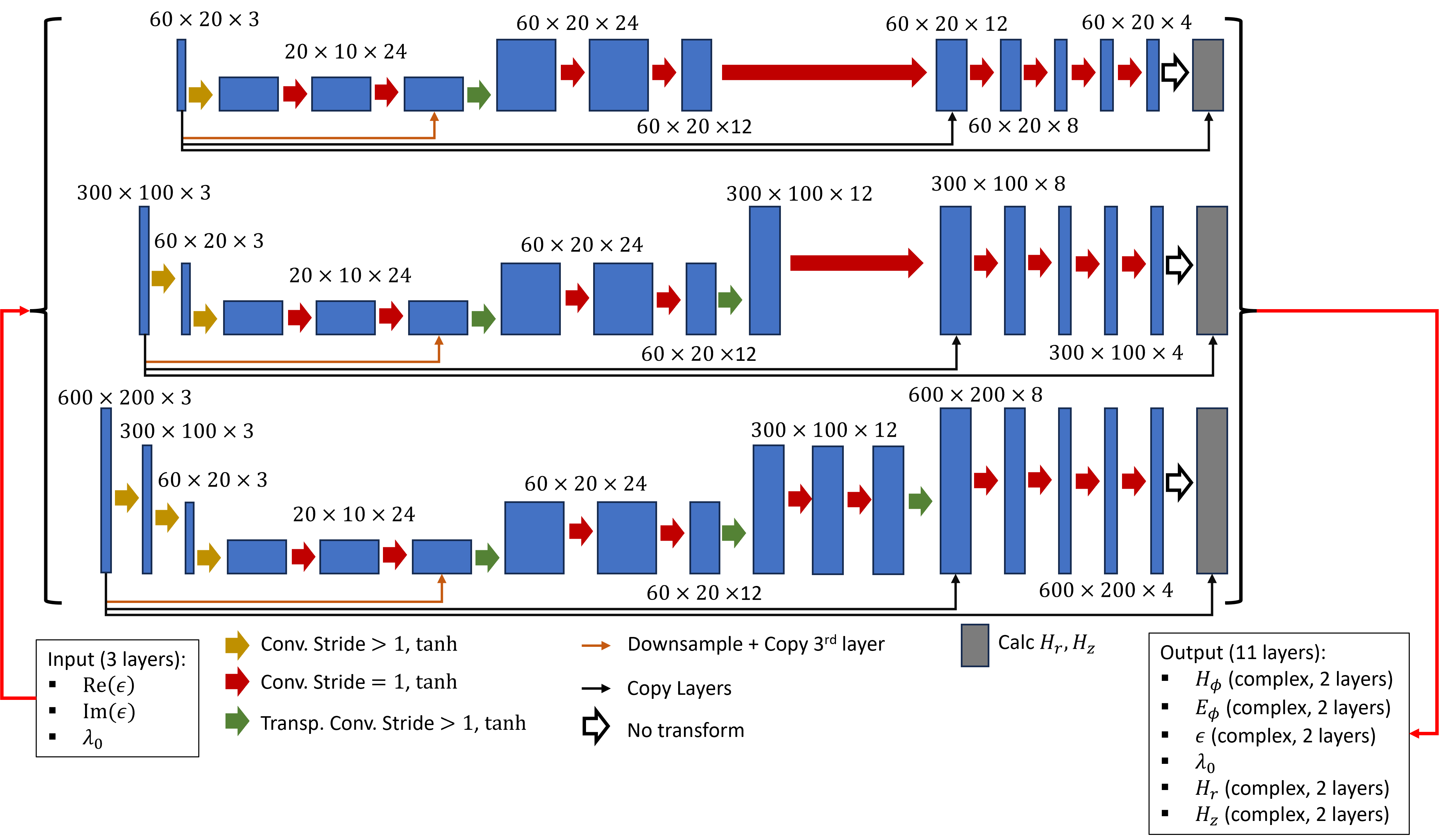}
		\caption{Setup of the CNN used in the study; the three rows represent low-, \mbox{medium-,} and high-resolution networks; boxes represent the size of data as it propagates through the network; arrows represent CNN data operations: each thick solid arrow represents the combination of a (transposed) convolutional layer and a $\tanh$ activation layer; thin black and orange arrows represent skip connections; thin red arrows represent input and output.  }
		\label{fig:architecture}
	\end{figure}

	We follow the general approach of constructing U-nets. The design of our networks is summarized in Fig.\ref{fig:architecture}. Starting with the pixel resolution of the data set, the proposed CNNs reduce the dimensionality of the problem to $20\times 10$ pixels, and then expand the resulting distributions to their original size. 
	
	The linear parts of the network employ standard convolutional and transposed convolutional layers with stride $=1$ for those parts of the network that preserve pixel size and with stride $>1$ for those that perform encoding/downsampling and decoding/upsampling. Hyperbolic tangent activation layers are used to add nonlinearities to the CNN. Combinations of convolutional and $\tanh$ layers are marked as thick arrows in Fig.\ref{fig:architecture}. In addition, custom layers are introduced to implement skip connections which can propagate the vacuum wavelength and permittivity distributions into the depth of the network for both stability of the resulting NN and to enable evaluation of  the physics-consistency of the resulting predictions. These layers operate by directly appending several layers of pixels to the output of a given convolutional layer (thin black arrows in Fig.\ref{fig:architecture}) or by first downsampling to the core resolution and then concatenating (thin orange arrows in Fig.\ref{fig:architecture}). 
	
	The base part of the NN (blue layers in Fig.\ref{fig:architecture}) is designed to learn the distribution of the $\phi$ components of the electric and magnetic fields. Note that in our hierarchical setup, the core of the networks remains the same, independent of the resolution of the data set, with the outer structure producing encoding/decoding from/to the higher resolution. The inner structure of the network (layer dimensionality and filter size) was optimized using the low-resolution data set. The medium- and high-resolution networks build upon this geometry by adding ``hierarchical'' downsampling and upsampling layers, implemented via convolutional and transposed convolutional layers in our NNs. Our analysis suggests that it is important to initialize the downsampling layers with unit weights, thereby setting the network for brute-force averaging of permittivity during the initial training iterations. 
	
	The physics-agnostic portion of the CNN, which is trained to produce the $\phi$ components of the magnetic and electric fields, is followed by a physics-informed layer (gray layers in Fig.\ref{fig:architecture}) which calculates distributions of the  $r$ and $z$ components of the magnetic field based on analytical expressions derived from Maxwell's equations: 
	\begin{eqnarray}
		\vec{E}_{rz} = \frac{-i}{\epsilon \frac{\omega^2}{c^2} - \frac{1}{r^2}}\left(-\frac{1}{r}\vec{D}_{rz} E_\phi - \frac{\omega}{c}\hat{\phi}\times\vec{D}_{rz}H_\phi\right) \\
		\vec{H}_{rz} = \frac{-i}{\epsilon \frac{\omega^2}{c^2} - \frac{1}{r^2}}\left(-\frac{1}{r}\vec{D}_{rz} H_\phi + \epsilon\frac{\omega}{c}\hat{\phi}\times\vec{D}_{rz}E_\phi\right)\label{eq:H}
	\end{eqnarray}
	where we have introduced the vector differential operator $\vec{D}_{rz}f = \hat{r}\frac{1}{r}\frac{\partial}{\partial r}\left(rf\right) + \hat{z}\frac{\partial f}{\partial z}$. 
	
	Because the fields are discretized on a regular rectangular grid, all derivatives are approximated with finite difference schemes. Forward and backward differences are used at the edges of the computational domain, while central differences are used within it. Our implementation of the CNNs used in this work and the data sets used in training are available on GitHub and Figshare, respectively\cite{PGMLgit,PGMLdata}.
	
	As seen from Eqs.(\ref{eq:H}), predictions for $H_r$ and $H_z$ may diverge when $\epsilon r^2 \omega^2/c^2 \approx 1$. This instability is a direct consequence of applying differential operators in cylindrical geometry. Here, we address the related issues by introducing a regularizing function (see below and SI). Our approach, illustrated here on an example of cylindrical geometry may be generalized to other curvilinear coordinates.
	
	\subsection{Knowledge transfer between different NN}
	As mentioned above, in the limit of ultrathin layers, the optics of multilayer metamaterials can be adequately described by the effective medium theory. In a related but separate scope, the U-shaped NNs are hypothesized to learn low-dimensional representations of the underlying phenomena. These considerations motivate the hierarchical design of the NNs used in this work. 
	
	To explore whether the learning outcomes of the NNs are consistent with the effective medium description, we performed a series of experiments where pretrained lower-resolution networks were used as pretrained cores of higher-resolution \emph{transfer-learning} (TL) networks. In these studies, the learning parameters of the pretrained ``core'' layers were frozen, with only the averaging and transposed convolution peripheral layers of the higher-resolution NN being trained. 
	
	At the implementation level, we drew inspiration from the ResNet \cite{7780459} architecture's approach of organizing layers into ``residual blocks.'' Specifically, we grouped the frozen layers into a single block, with the internal layer weights corresponding to those of the selected pretrained network. The forward function was designed to perform training within the layers of the block; however, during backpropagation, the weight updates bypass the internal layers of the block, passing directly to the previous layer. 
	
	We explored knowledge transfer from low- to medium-resolution networks as well as from medium- to high-resolution networks.

	\subsection{Training protocols}
	
	\sloppy To assess the benefits of the physics-based constraints, three different regimes of training the CNN are explored. In the base-case \emph{black-box} (BB) scenario, the model minimizes only the radially-weighted mean-squared error of the $\phi$ components of the electric and magnetic fields (directly produced by the physics-agnostic part of the network) ${L_{\phi}= \left\langle w(r)\left[\left|H_\phi^Y - H_\phi^T\right|^2 + \left|E_\phi^Y - E_\phi^T\right|^2\right]\right\rangle}$. Here, the superscripts $Y$ and $T$ correspond to the predicted and ground-truth fields, respectively,  the angled brackets, $\langle \cdots \rangle$, represent an arithmetic mean over the simulation region, and  the radial weight function, $w(r)$, is used to emphasize the region of small radii where the funnel is located. 
		
	The second, \emph{field-enhanced} (FE) model utilizes a hybrid loss that combines the above-described $L_\phi$ with its analog for the remaining components of the magnetic field, 
	\begin{equation}
		L_{FE}= L_{\phi} + L_{rz} 
	\end{equation}
	with  ${L_{rz}= \left\langle w(r)\left|R\right|^2\left[\left|H_r^Y - H_r^T\right|^2 + \left|H_z^Y - H_z^T\right|^2\right]\right\rangle}$ and the $rz$ components of the magnetic field being produced by the physics layer of the CNN. 
	
	In order to prevent the instability of Eq.(\ref{eq:H}) from dominating the overall loss, we introduce the regularization function, $R(r,z)$, such that $R(r,z)\rightarrow 0$ when $r\rightarrow c/(\sqrt{\epsilon(r,z)}\omega)$ (see the supplementary information for details). 
	
	Because calculation of the $r$ and $z$ field components requires differentiating the $\phi$ components, the addition of $L_{rz}$ allows the CNN to learn the relationships between the spatial field distributions and the distributions of their derivatives. Importantly, evaluation of both $L_\phi$ and $L_{rz}$ terms requires the training set to contain the solutions of Maxwell's equations (labeled data).

	Finally, \emph{physics-guided} (PG) training combines the above labeled-data--dependent terms, $L_\phi$ and $L_{rz}$, with the physics loss, 
	\begin{eqnarray}
		L_{ph}= \frac{1}{\max \left| H_\phi^Y \right|}\left\langle\left|\frac{\partial(H_z^Y R^2)}{\partial r} - \frac{\partial(H_r^Y R^2)}{\partial z} + i\frac{\omega}{c}\epsilon\ E_\phi^Y R^2 - 2\left(H_z^Y R \frac{\partial R}{\partial r} - H_r^Y R \frac{\partial R}{\partial z}\right)\right|\right\rangle
		\label{eq:Lph}
	\end{eqnarray}
	which represents the (regularized) residual of Maxwell's equations for the $H_\phi$ component of the field (see the supporting information). Therefore, PG training  aims to enforce consistency of the solutions that are generated by the NN with Maxwell's equations.
	Notably, evaluation of the physics loss does not require labeled data. As a result, \emph{unlabeled-trained} (UL) networks can utilize a combination of labeled and unlabeled data, with the former inherently incorporating the boundary conditions, and the latter allowing the expansion of the training set without computing additional PDE solutions. This UL loss was also used in training the TL networks described in the preceding section
	
	Previous analysis\cite{eager-1} demonstrated that BB- and PG-loss often compete with each other. Here, this competition reflects the different differentiation schemes used by FEM and the PG-loss as well as the existence of multiple solutions to Maxwell's equations (for example, the trivial  solution $\vec{E}=\vec{H}=0$) that do not necessarily satisfy the boundary conditions that are implicitly enforced by labeled data. To guide the network towards the correct implementation of boundary conditions, the weight of the physics-loss, $w_{ph}$, is dynamically adjusted during training\cite{eager-1}, resulting in the dynamic PG loss,
	\begin{eqnarray}
		L_{PG}= L_{\phi}+ L_{rz}+ w_{ph}L_{ph}\text{.}
	\end{eqnarray}
	
	In order to assess the ability of the networks to interpolate and extrapolate between data sets having plasmonic layers with different plasma wavelengths, we train the networks on 50\% of the data with plasma wavelengths of 6 and \SI{11}{\micro\meter} \emph{or} with plasma wavelengths of 7 and \SI{10}{\micro\meter}, and add up to 10\% of the labeled data from other data sets to the training. The UL models are also provided configurations from the remaining data sets as unlabeled data. The training scenarios are summarized in Table \ref{tab:datasets}, which gives the percent of each data set that was used as labeled and unlabeled data in each network type. Each training scenario has been used to train at least 10 different networks of each resolution  and loss type, with the dynamics of their training and validation loss presented in the supplementary information, and their averaged performance summarized below. 
	
	\begin{table}
		\begin{center}
			\caption{Labeled and Unlabeled Training Data Composition of Each Network}
			\label{tab:datasets}
			\begin{tabular}{l |l l l l l| l l l l l }
				& \multicolumn{5}{c|}{Labeled \% } & \multicolumn{5}{c}{Unlabeled \%} \\
				\multicolumn{1}{c|}{Network} & \multicolumn{5}{c|}{$\lambda_p$ (\unit{\micro\meter})} & \multicolumn{5}{c}{$\lambda_p$ (\unit{\micro\meter})} \\
				& 6& 7& 8.5& 10& 11 & 6& 7& 8.5& 10& 11  \\
				\hline
				BB$_i$, FE$_i$, PG$_i$ & 50 & 0& 10& 0& 50& \multicolumn{5}{c}{none}\\ 
				UL$_i$, TL$_i$ & 50 & 0& 10& 0& 50 & 0 & 0& 40 & 0& 0 \\ 
				BB$_e$, FE$_e$, PG$_e$ & 0 & 50& 10 & 50& 0& \multicolumn{5}{c}{none}\\ 
				UL$_e$ & 0 & 50& 10 & 50& 0 & 25 & 0& 40 & 0& 25\\ 
				BB$_x$, FE$_x$, PG$_x$ & 10 & 50& 10 & 50& 10& \multicolumn{5}{c}{none}\\ 
				UL$_x$ & 10 & 50& 10 & 50& 10 & 25 & 0& 40 & 0& 25\\
			\end{tabular}
		\end{center}
	\end{table}

	\section{Results}
	
	To demonstrate the impact of physics-based constraints on the accuracy and consistency of NN-predicted fields, we analyze the dependence of  the three average losses introduced above ($L_\phi$, $L_{rz}$, and $L_{ph}$) both on the enforcement of physics-consistency and on the presence of unlabeled data during training. Sample field distributions are presented to illustrate the models' performance. Finally, we analyze the generalizability of the models by evaluating their performance across the plasma wavelengths of the plasmonic layers.
	
	The three components of the loss, $L_\phi, L_{rz}$ and $L_{ph}$, are arranged in increasing degree of physics consistency and -- simultaneously -- decreasing reliance on data. Indeed, $L_\phi$, which analyzes only the physics-agnostic output of the networks, relies exclusively on data. $L_{rz}$, which primarily relies on the output of the physics layer enforces the relationships between the fields at neighboring points [see Eq.(\ref{eq:H})]. Lastly, $L_{ph}$ exclusively analyzes physics-consistency and pays no regard to data consistency. Our analysis (see below) illustrates that training with $L_{ph}$ not only improves the consistency with Eq.(\ref{eq:Lph}) but also improves other metrics that are related to Maxwell's equations, such as energy conservation -- as analyzed through the Poynting theorem (see SI).

	\begin{figure}[h!]
		\centering
		
		\includegraphics[width=\linewidth]{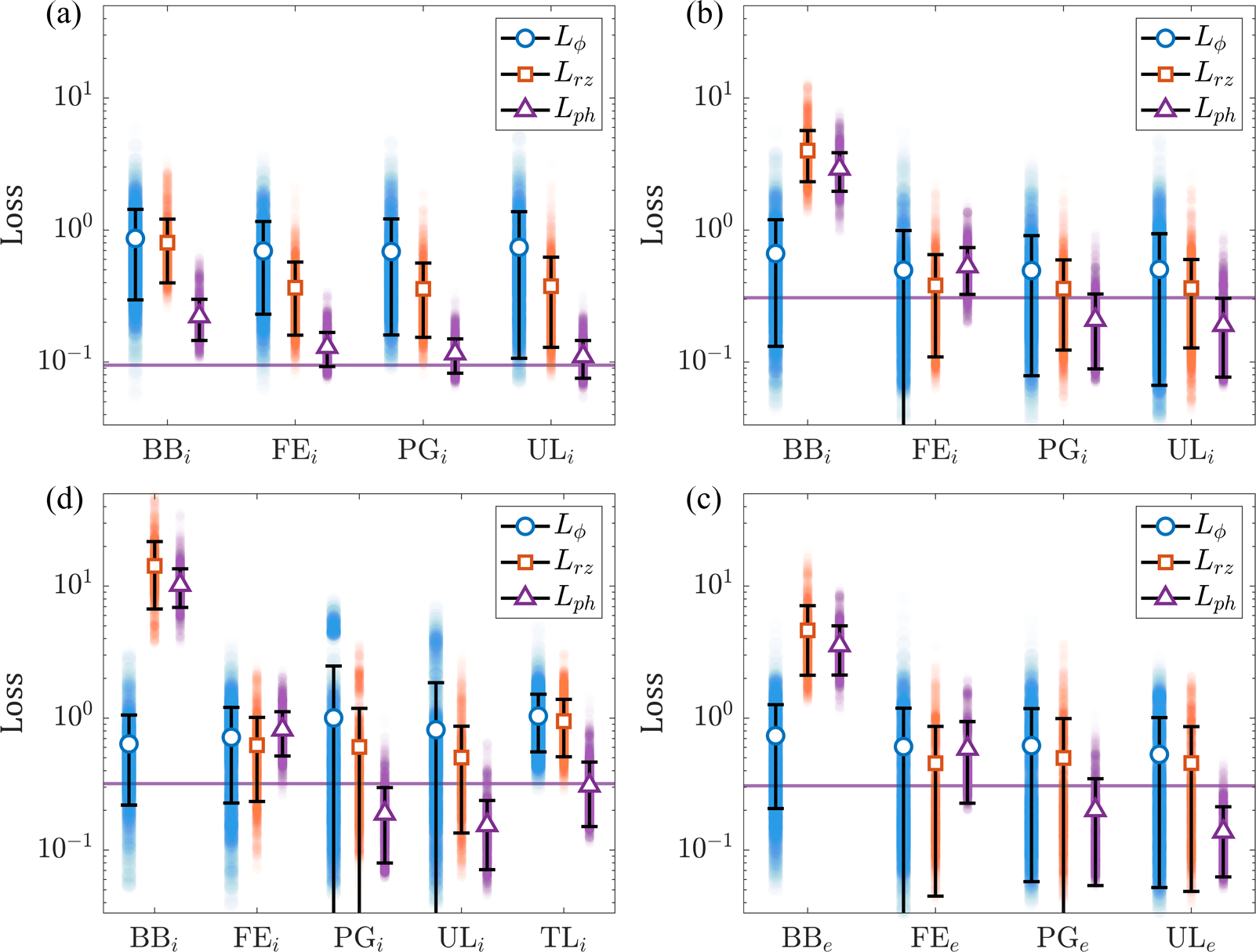}
		\caption{Performance of NNs with different architectures and training protocols, evaluated on the data that was not used in training; panels (a,b,c,d) represent low-resolution (a), medium resolution (b,c), and high-resolution (d) networks (see Table 1 for network labels); loss metrics of individual predictions are represented as filled semitransparent circles, resulting in the color-coded distributions; solid white markers and black bars represent the mean and standard deviation of these distributions; the purple horizontal lines show the average $L_{ph}$ of all interpolated FEM solutions.}
		\label{fig:avgmodel}
	\end{figure}
	
	\subsection{Impact of Physics Information on Accuracy}
	
	The performance of the different models is summarized in Fig.\ref{fig:avgmodel}.  With the comparatively simple low-resolution model, adding the physics-based layer to the network and adding the $L_{rz}$ component to the loss function provides enough additional information to adequately represent the coarsely sampled data. Providing the network additional physics-based information (by implementing PG loss) does not quantitatively boost the performance of the model -- due to a combination of the model's simplicity and the  mesh being too coarse to resolve the composite structure. 
	
	As the resolution and complexity of the model grow, increasing physics-based constraints and adding unlabeled data yields measurable improvements in model performance. Interestingly, the extra physics consistency (as demonstrated by the improving $L_{ph}$ metric) sometimes comes at the cost of a small increase of $L_\phi$. This apparent contradiction results from the fact that the data used in training was generated by reinterpolating FEM solutions from a triangular mesh to a rectangular mesh. As a result, the ``ground truth'' does not yield vanishing $L_{ph}$. As seen in Fig.\ref{fig:avgmodel}, predictions of the neural net tend to be closer solutions to Maxwell's equations on the rectangular mesh than the FEM-sourced data. 
	
\begin{figure}
		\centering
		\includegraphics[width=\linewidth]{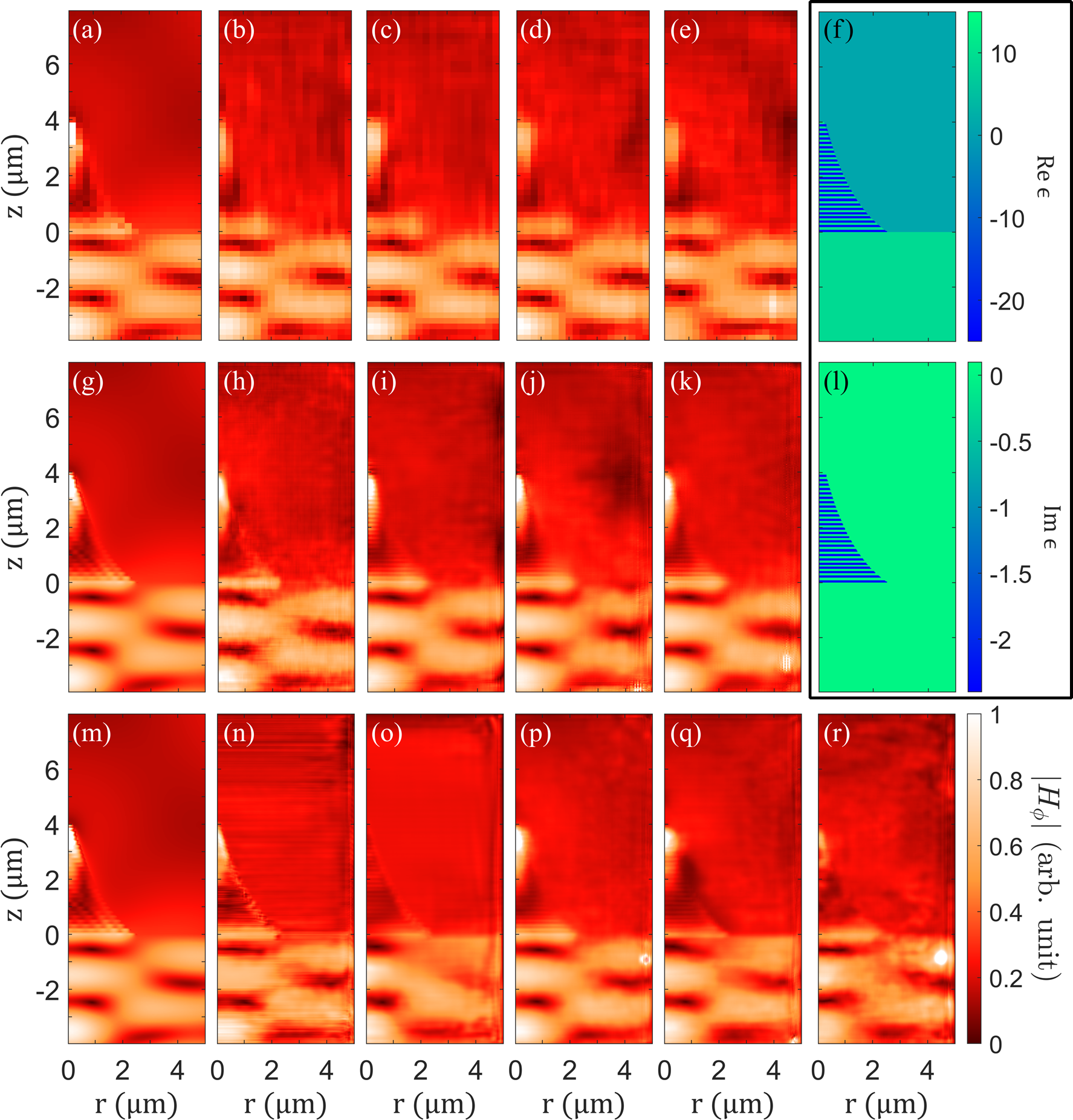}
		\caption{Representative predictions of the NNs with (a..e) low- , (g...k) medium-, and (m...r) high-resolution; input permittivity is shown in panels (f,l); panels (a, g, m) represent ground truth; panels (b, h, n) - predictions of BB$_i$ NNs, panels (c, i, o) - predictions of FE$_i$ networks, panels (d, j, p) - predictions of PG$_i$ networks, and panels (e, k, q) - predictions of UL$_i$ networks; panel (r) illustrates the performance of the TL$_i$ network. Note that higher-performing networks resolve field oscillations on the scale of individual layers within the composite and field concentration near the tip of the funnel.
		}
		\label{fig:fields}
	\end{figure}

	A more granular look at the NN predictions is shown in Fig.\ref{fig:fields} where representative examples of model predictions are compared with FEM solutions. Note that in contrast to their BB counterparts, PG networks predict smoother fields and resolve individual layers of the structure.  
	
	Our results are in agreement with previous studies\cite{zhelyeznyakov2023large,chen2022high} that focused on predictions of field distributions in dielectric structures trained on relatively large ($\sim 10^4$ configurations) data sets. Incorporation of physics loss in these NNs resulted in substantial (but limited) improvements in physics consistency (by a factor of $\lesssim 2$). Here, we see similar dynamics for low-resolution networks that require few labeled-data training inputs to achieve their top performance. At the same time, the physics consistency of our medium- and high-resolution networks, which are trained in the data-poor regime, is improved by an order of magnitude as a result of the incorporation of physics-based constraints.

	\subsection{Knowledge Transfer}
	
	As mentioned above, we have attempted knowledge transfer from a pretrained low-resolution network to a medium resolution network and from a pretrained medium resolution network to its high-resolution counterpart. In both cases, a single average-performing lower-resolution UL network was chosen as the source of the frozen core of the higher-resolution TL networks. Notably, the low-resolution network poorly resolves the individual layers within the composite. Consistent with this design, implementation of PG loss does not substantially improve network performance (see above), and using a pretrained low-resolution network as a learning-free core of the medium-resolution counterpart does not yield adequate performance of the resulting NN. 
	
	In contrast, using a pretrained medium resolution network as a (fixed) core of a high-resolution NN provided reasonable performance. As seen in Fig.\ref{fig:avgmodel} and Fig.\ref{fig:fields}, the accuracy of TL$_i$ networks falls between the fully-trained high-resolution FE$_i$ and PG$_i$ NNs.  
	
	The physics of finely stratified composites is {\it analytically} described by effective medium theories (EMT). In the EMT formalism, the spatial distribution of homogenized (averaged over the scale of the inclusion $\sim d$) electromagnetic fields is given by effective parameters (here, $\epsilon_\perp$ and $\epsilon_{zz}$). These homogenized fields, along with equations that relate the effective medium parameters to microscopic distributions of permittivity, can then be used to recover fine-scale field distributions.\cite{vpTutorialsBook}
	
	The analytical procedure described above is somewhat similar to the operation of the hierarchical TL CNN reported in this work. Indeed, the CNN-based U-nets are known to learn a low-dimensional representation of the underlying phenomena. From this standpoint, while we do not analyze the neural operation of the CNN in detail, the medium-resolution network is likely to learn some form of materials averaging/field recovery by analyzing the transition between the scale of individual layers (resolved at the entrance and exit of the network) and compact representations in its core. The transfer-learning high-resolution wraparound parts of the network likely learn the averaging and upscaling procedures. We reserve the analysis of the relationship between the analytical EMT and the operation of TL-based hierarchial CNNs for future work.

	By freezing the inner core of the CNN within knowledge transfer networks we significantly reduce the number of training parameters. Therefore, we expect smaller variability and faster learning in the TL$_i$ networks as compared with their fully-trained high resolution PG$_i$ counterparts. However, in our implementation, the time required to calculate one training epoch of a TL$_i$ network is almost identical to the time required for one epoch of a PG$_i$ network, indicating that the time spent updating the learnable NN parameters is significantly less than the time spent executing forward and backward propagation steps. Different implementation and optimization settings may affect this result.
	
	At the same time, further analysis (see SI) suggests that TL$_i$ networks converge over a smaller number of epochs than their PG$_i$ counterparts. In addition, in our studies, variation between the performance of the best and the worst TL$_i$ networks was significantly smaller than the variation between the best and the worst PG$_i$ networks.
	
	\subsection{Interpolation vs Extrapolation within the models}

	As described above (see Table \ref{tab:datasets}), the NNs have been trained on multiple subsets of the data derived from FEM solutions, aiming to assess both correctness and generalizability of the proposed PGML networks. Here we are particularly interested in the ability of the NN to generalize the results between different plasma wavelengths of the doped components of the funnels' cores.  
	
	In the ``interpolating'' models (subscripted $i$), 50\% of the data from the sets representing the lowest and the highest plasma frequencies, and an additional 10\% from the data set representing the central plasma wavelength were used as labeled training data. The unlabeled networks further included 40\% of the central plasma wavelength data set as unlabeled data. Therefore, the CNN would have to deduce the behavior of the composites with $\lambda_p=7,10\,\si{\micro\meter}$.  For the ``extrapolating'' (subscripted $e$) and ``extended extrapolating'' (subscripted $x$) networks, a similar approach was used except with the bulk of labeled training data coming from the 7 and 10 \unit{\micro\meter} plasma wavelengths, having the CNN deduce the behavior of the metamaterials with $\lambda_p=6,11\,\si{\micro\meter}$

	Typically, data interpolation is a much simpler problem than data extrapolation. However, this general rule does not hold for our analysis. 
	As seen in Fig.\ref{fig:avgmodel}(b,c), the average performance of the two classes of medium-resolution networks is almost identical to each other, indicating that both interpolation and extrapolation tasks (in terms of $\lambda_p$) in our study represent similar difficulties to the NNs.
	
\begin{figure}[H]
		\centering
		\includegraphics[width=\textwidth]{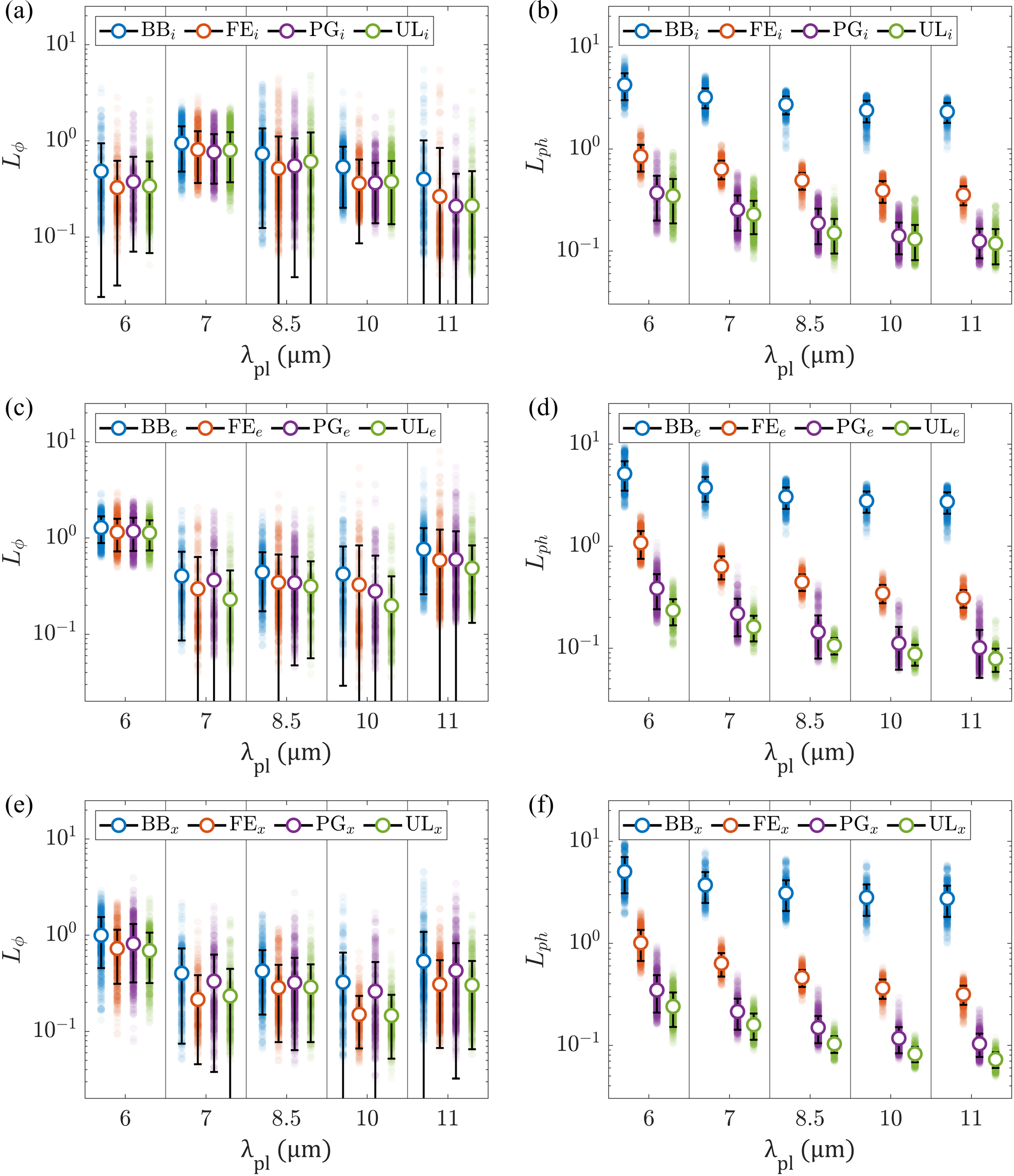}
		\caption{Performance of the medium-resolution networks for predicting the field distribution of composites with given plasma wavelengths; panels (a, c, e) and (b, d, f) represent $L_\phi$ and $L_{ph}$, respectively, for (a, b) interpolating, (c, d) extrapolating, and (e, f) extended extrapolating networks; colors represent training protocols; individual predictions are represented as filled semitransparent circles, resulting in the color-coded distributions; solid white circle markers and black bars represent the mean and standard deviations of these distributions.}
		\label{fig:midplasmas}
	\end{figure}

	Fig. \ref{fig:midplasmas} provides a more in-depth look at this behavior. In general, as characterized by $L_\phi$ loss, the networks perform their best in predicting the fields within the metamaterials for the same plasma wavelength that comprises the majority of their labeled training set. Indeed, $L_\phi$ is $\sim 2$ times lower for the data that has a plasma wavelength that is well-represented in the training set than for the configurations with plasma wavelengths that contribute few or no instances to the labeled training data. Incorporation of physics-based constraints improves the physics-consistency of the results for all values of $\lambda_p$ by an order of magnitude, indicating that the CNNs learn the general properties of the field distributions but miss the particular boundary conditions that are encoded in the labeled data.
	
	By comparing the performance of extrapolating networks to their ``extended'' counterparts [Fig.\ref{fig:midplasmas}(c,e)] it is seen that adding very little labeled data can somewhat address this issue of underspecified boundary conditions: introduction of $\sim 20$ labeled distributions (total) for $\lambda_p=6,11\mu m$ reduces the $\lambda_p$-specific $L_\phi$ by $\sim 20\%$ with almost no effect on $L_{ph}$.
	
	Interestingly, in all scenarios $L_{ph}$ decreases as a function of $\lambda_p$. This behavior traces the strength of the resonance in $\epsilon_{zz}$ that decays and moves out of the spectral range of the study as $\lambda_p$ increases (see Fig.\ref{fig:schematic}d).

	\section{Conclusions}

	We have presented a hierarchical design of PG neural network surrogate solvers of Maxwell's equations and demonstrated the proposed formalism by predicting the field distributions in hyperbolic metamaterial-based photonic funnels. We have demonstrated that embedding physics information into the ML process, by  enforcing the physics-based constraints and by adding unlabeled training configurations, improves the quality of ML predictions in the regime of limited training data. In particular, physics-guided ML predictions are almost two orders of magnitude more physics-consistent than their BB-ML counterparts, even near wavelengths where the layered composite undergoes topological transitions. Separately, we have demonstrated that a hierarchical network architecture enables knowledge transfer from existing pretrained models to higher-resolution NN implementations.  
	
	The approach presented can be directly applied to the analysis of complex rotationally-symmetric electromagnetic systems. The technique can be straightforwardly extended to quasi-2D geometries with inclusions of various sizes and shapes by using the appropriate coordinate-representations of Maxwell's equations. The formalism can be further extended to 3D geometries, although we anticipate that such extensions will require significantly larger computational resources.

	\section*{Funding Sources}
	The work has been supported by the National Science Foundation (awards\# 2004298, 2423215, 2004422, 2026710, and 2239328).

	\bibliography{refs}
	
	\newpage
	\appendix
\section*{Supplemental material}

\renewcommand{\theequation}{S\arabic{equation}}
\renewcommand{\thefigure}{S\arabic{figure}}
\setcounter{equation}{0}
\renewcommand\thesubsection{S\arabic{subsection}}

\subsection{Maxwell's Equations in Cylindrical Geometry}
As mentioned in the main manuscript, we use the rotational symmetry of our problem to reduce the three-dimensional vectorial Maxwell's equations to equations describing the behavior of the $\phi$-components of the electric and magnetic fields (which vary smoothly throughout the geometry). Assuming that all fields are proportional to $\exp(-i\phi)$, once $E_\phi$ and $H_\phi$ are known (for example, as predicted by the neural net), the remaining components of the fields can be calculated via 
\begin{eqnarray} 
    \vec{E}_{rz} = \frac{-i}{\epsilon \frac{\omega^2}{c^2} - \frac{1}{r^2}}\left(-\frac{1}{r}\vec{D}_{rz} E_\phi - \frac{\omega}{c}\hat{\phi}\times\vec{D}_{rz}H_\phi\right) \label{eq:E}\\
    \vec{H}_{rz} = \frac{-i}{\epsilon \frac{\omega^2}{c^2} - \frac{1}{r^2}}\left(-\frac{1}{r}\vec{D}_{rz} H_\phi + \epsilon\frac{\omega}{c}\hat{\phi}\times\vec{D}_{rz}E_\phi\right)\label{eq:supH}
\end{eqnarray}
where the differential operator $\vec{D}_{rz}$ is defined by $\vec{D}_{rz}f = \hat{r}\frac{1}{r}\frac{\partial}{\partial r}\left(rf\right) + \hat{z}\frac{\partial f}{\partial z}$. 

Maxwell's equations also provide additional constraints, ensuring self-consistency of the field components: 
\begin{eqnarray}  
    \frac{\partial {H}_r}{\partial z} 
    -\frac{\partial {H}_z}{\partial r} 
    = -i\epsilon\frac{\omega}{c}{{E}_\phi} \label{eq:supEphi}
\\    
    \frac{\partial {E}_r}{\partial z} 
    -\frac{\partial {E}_z}{\partial r} 
    = i\frac{\omega}{c}{{H}_\phi}
\end{eqnarray}
the first of which is used as a basis for $L_{ph}$ in the manuscript.

\subsection{Regularization Function}

The neural network directly predicts $E_\phi$ and $H_\phi$, and the physics layer then calculates $H_r$ and $H_z$ using Eq.(\ref{eq:supH}), above. It is seen, however, that for transparent materials, Eq.(\ref{eq:supH}) diverges when $r^2 \epsilon \omega^2/c^2=1$. 

In approximate numerical solutions (such as those analyzed in our work), this condition leads to instabilities that -- if left unaddressed -- would dominate both $L_{rz}$ and $L_{ph}$ loss functions. 

To address these underlying instabilities, we introduce the regularization function $R(r,z)$ in such a way that regularized fields $\vec{\mathcal{E}} = R\vec{E}$ and $\vec{\mathcal{H}} = R\vec{H}$ remain finite within the simulation domain. Explicitly, 
\begin{equation}
    R(r,z) = 0.1\frac{\epsilon(r,z) r^2 \frac{\omega^2}{c^2}-1}
    {\epsilon(r,z) r^2 \frac{\omega^2}{c^2} + 0.1}
\end{equation}
is used in our work (In principle, any $R(r,z)$ that vanishes, at least linearly, when $r^2 \epsilon \omega^2/c^2=1$ can be used).

 To find an appropriate physics loss function, we first recast Eq.(\ref{eq:supEphi}) in terms of regularized fields
\begin{equation}
    \frac{\partial}{\partial z} \left(\frac{\mathcal{H}_r}{R}\right)
    -\frac{\partial}{\partial r} \left(\frac{\mathcal{H}_z}{R}\right)
    = -i\epsilon\frac{\omega}{c}\frac{\mathcal{E}_\phi}{R}.
\end{equation}
We then apply the derivatives and rearrange the resulting relationships, arriving at: 
\begin{equation}
    \frac{\partial}{\partial z}\left(R\mathcal{H}_r\right) 
    - \frac{\partial}{\partial r}\left(R\mathcal{H}_z\right)
    +2\left(
        \mathcal{H}_z\frac{\partial R}{\partial r}
        -\mathcal{H}_r\frac{\partial R}{\partial z}
    \right)
    +i\epsilon\frac{\omega}{c}R\mathcal{E}_\phi = 0. 
\end{equation}
Recasting the latter equation back to the actual fields yields the physics residual used for physics loss in our work
\begin{equation}
    \Lambda_{ph}=\frac{\partial}{\partial z}\left(R^2 H_r\right) 
    - \frac{\partial}{\partial r}\left(R^2 H_z\right)
    +2\left(
        R H_z\frac{\partial R}{\partial r}
        -R H_r\frac{\partial R}{\partial z}
    \right)
    +i\epsilon\frac{\omega}{c}R^2 E_\phi \text{.}
\end{equation}
Note that physics-consistent solutions should satisfy $\Lambda_{ph}(r,z)\equiv 0$.

\subsection{Radial Weight Function}
The fundamental electromagnetic phenomena enabled by photonic funnels (anomalous reflection and subdiffractive light confinement) are encoded in the field distributions within and in close proximity to the funnels. Since these field distribution features are more important for understanding the electromagnetism of the funnels and at the same time are more complicated than the diffraction-limited field distributions outside the funnels, a weight function is used to improve learning of the fields at small radii. In our networks, this radial weight function is a sigmoid given by
\begin{equation}
    w(r) = \frac{5}{1+10 e^{2(r-3)}} + 0.5\text{,}
\end{equation}
where $r$ is in \si{\um}.

\subsection{Physics-Consistency and Energy Conservation}
An important physical principle, which should hold (at least approximately, given our discretization) for physically consistent fields is energy conservation. We may therefore use the deviation from energy conservation as an additional measure of physics-inconsistency. Furthermore, we can utilize this metric to demonstrate the impact of imposing physics-consistency through the inclusion of $L_{ph}$ in the training loss function. 

\begin{figure}[h!]
	\centering
	
	\includegraphics[width=\linewidth]{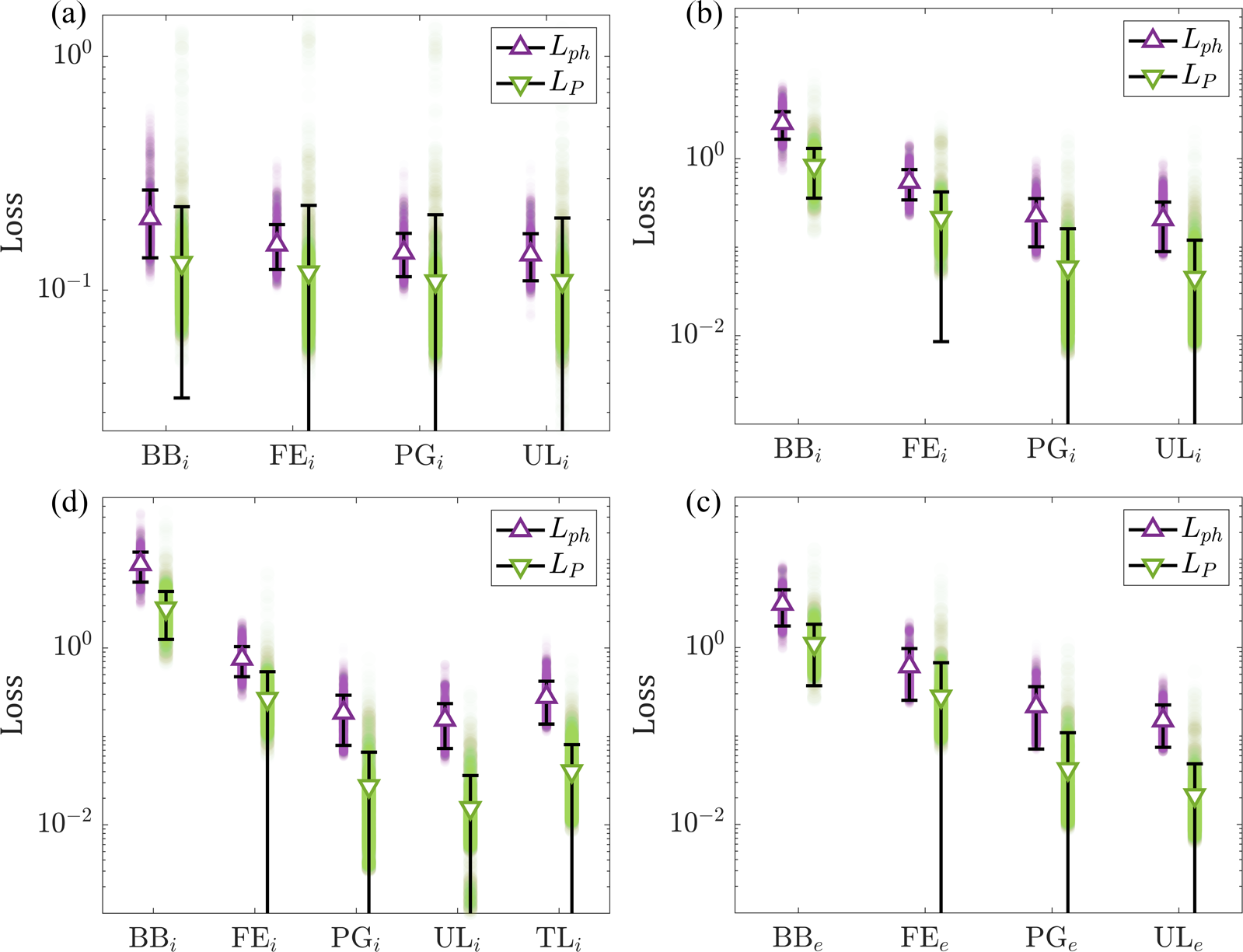}
	\caption{Physics and Poynting losses of NNs with different architectures and training protocols, evaluated on the subset of data that was not used in training sets; panels (a,b,c,d) represent low-resolution (a), medium resolution (b,c), and high-resolution (d) networks; Losses of individual predictions are represented as filled semi-transparent circles; solid white markers and black bars represent the mean and standard deviations of these distributions.}
	\label{fig:poyntingloss}
\end{figure}

The conservation of energy for monochromatic fields in linear dispersive media can be written as
\begin{equation}\label{eq:poynting}
    \frac{\omega}{c}\mathrm{Im}\left\{\epsilon\left|\vec{E}\right|^2+\mu\left|\vec{H}\right|^2\right\}-\nabla\cdot\mathrm{Re}\{\vec{E}\times\vec{H}^*\}=0\mathrm{.}
\end{equation}
The left-hand side of this equation acts as a kind of ``Poynting residual'' when applied to electromagnetic fields, with nonzero values representing local violations of energy conservation. 

As was the case with $\Lambda_{ph}$, to make use of this residual with network-predicted fields, we must suppress the numerical instabilities which arise from calculating the $r$ and $z$ field components via Eq.\ref{eq:E} and Eq.\ref{eq:supH}.
Because this residual expression is proportional to the square of the fields, adequately suppressing the numerical errors introduced in calculating these additional components requires multiplication by $R^2$ resulting  in the  regularized Poynting residual: 
\begin{equation}
    \Lambda_P = R^2\left(\frac{\omega}{c}\mathrm{Im}\left\{\epsilon\right\}\left|\vec{E}\right|^2-\nabla\cdot\mathrm{Re}\{\vec{E}\times\vec{H}^*\}\right)
\end{equation}
for non-magnetic materials. In analogy with how our physics loss is defined in terms of our physics residual, a Poynting loss, $L_P$, was calculated as the average magnitude of the Poynting residual normalized by the square of the maximum field,  $\max (|E_\phi|,|H_\phi|)$ (here, Gaussian units are used for convenience, making amplitudes of electric and magnetic fields of the plane wave comparable to each other).

To demonstrate that the physics-consistency enforced by $L_{ph}$ results in fields which better satisfy energy conservation, we calculated the Poynting loss for the predictions of networks of each size across all test data. The results are shown in Fig.\ref{fig:poyntingloss}. It is clearly seen that improving physics-consistency (characterized by lowering $L_{ph}$) yields  improvement in energy conservation (characterized by lowering $L_P$). 

\subsection{Loss Dynamics}

\begin{figure}[h!]
     \centering
     
  \includegraphics[width=\linewidth]{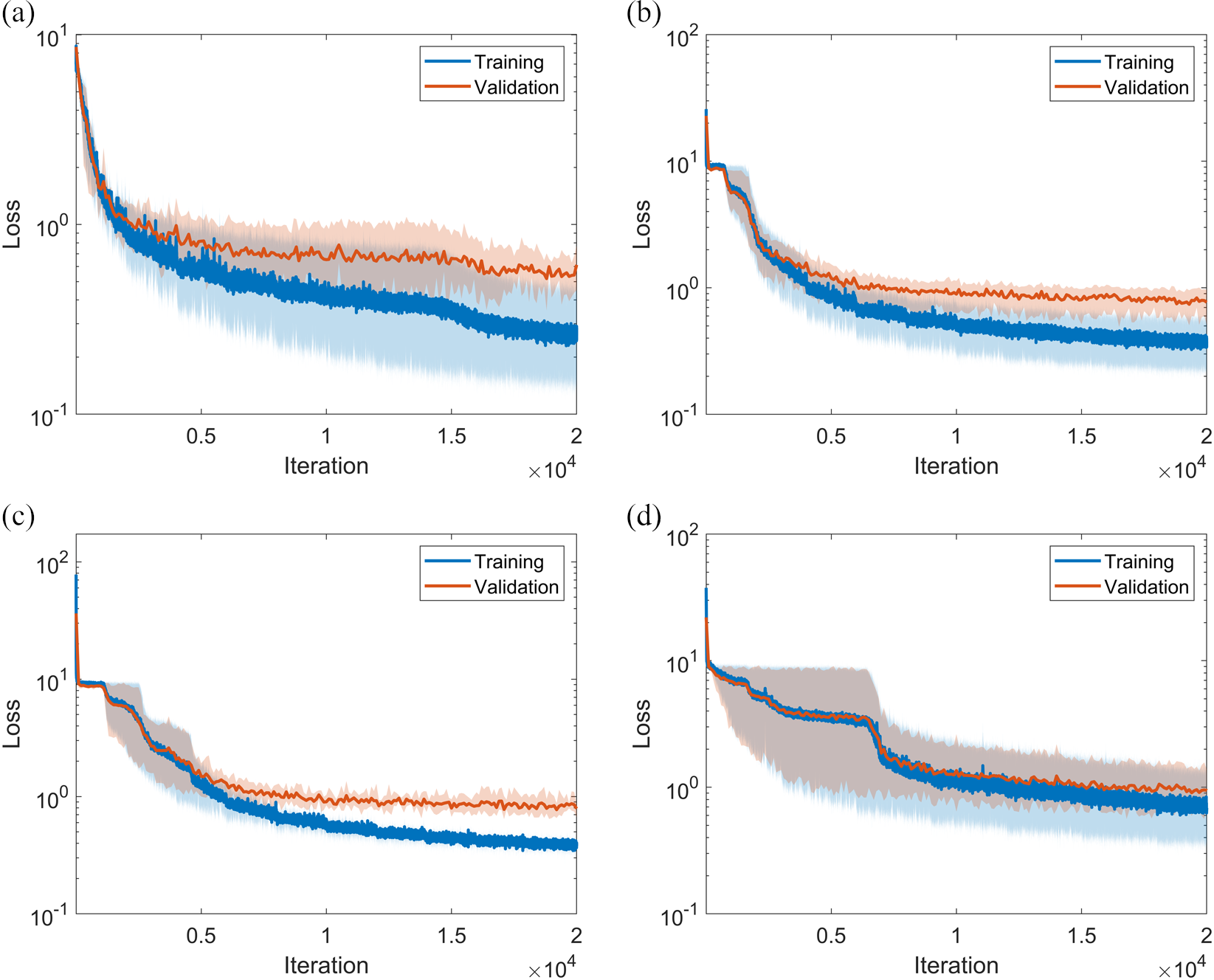}
        \caption{Loss dynamics of medium resolution networks showing training and validation loss against training iteration for (a) black-box, (b) field-enhanced, (c) physics-guided, and (d) unlabeled-trained networks. The solid lines show the losses averaged over networks while the shaded regions are bounded by the maximum and minimum losses at each iteration.}
        \label{fig:lossdynamics}
\end{figure}
\begin{figure}[h!]
	\centering
	
	\includegraphics[width=0.5\linewidth]{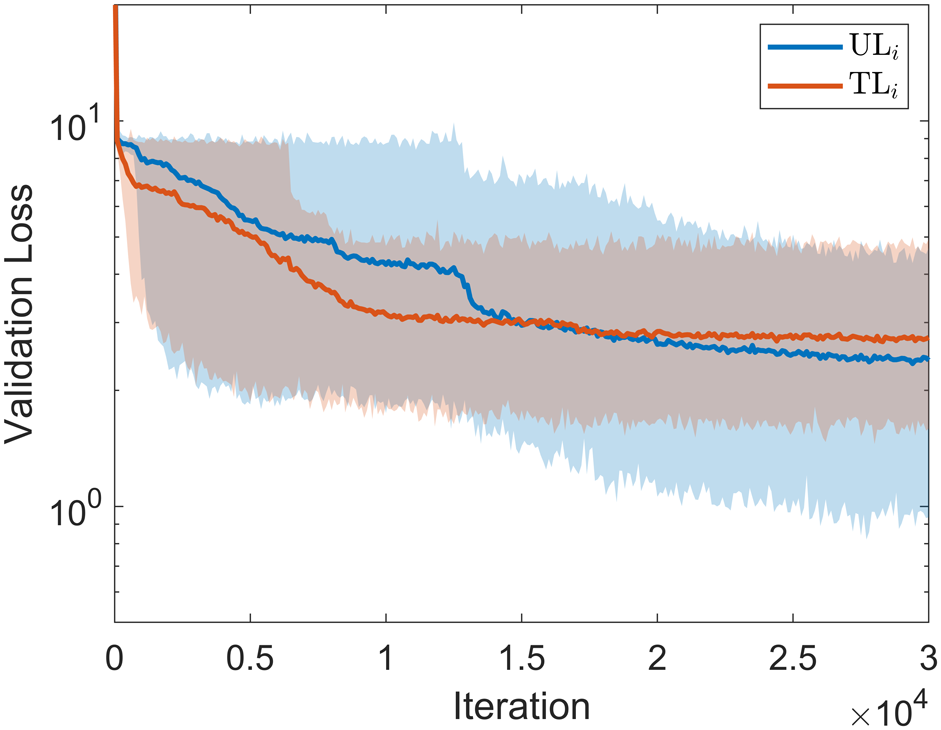}
	\caption{Validation loss dynamics of high-resolution unlabeled-trained and transfer-learning networks. The solid lines show the losses averaged over networks while the shaded regions are bounded by the maximum and minimum losses at each iteration.}
	\label{fig:ULvsTL}
\end{figure}

The training and validation loss curves for each of the four medium resolution ``interpolating'' network configurations are presented in Fig. \ref{fig:lossdynamics}. Across all cases, the validation loss closely follows the trend of the training loss throughout the training process. While the two losses begin at similar values---occasionally with the validation loss slightly lower during the early epochs---they gradually separate as training progresses with the validation loss becoming slightly greater towards convergence. This pattern is expected given the model's exposure to the training data and indicates stable, consistent generalization to unseen data, with no evidence of significant overfitting.

Similar results were seen for high-resolution networks, pointing again to stable training without overfitting. Additionally, we compared the validation loss dynamics between high-resolution UL$_i$ and TL$_i$ networks, summarized in   Fig.\ref{fig:ULvsTL}. It can be seen that TL$_i$ networks require fewer training epochs to converge and tend to outperform their UL$_i$ counterparts during early training iterations. However,  UL$_i$ networks tend to eventually outperform their TL$_{i}$ counterparts. 

This dynamics reflects the comparatively smaller parameter space of TL networks, which are therefore more readily able to find optimal network configurations and less likely to fall into (and become stuck within) local minima. However, this smaller network dimensionality comes at the price of expressiveness, somewhat limiting the ability of TL networks to fine-tune their predictions.

\end{document}